\documentclass[journal=ancac3,manuscript=article]{achemso}

\usepackage{amsmath}
\usepackage{amssymb}
\usepackage{dcolumn}
\usepackage{graphicx}
\usepackage[caption=false, label font=bf,labelformat=simple]{subfig}

\usepackage{siunitx} 
\usepackage{physics} 
\usepackage[normalem]{ulem}
\usepackage{soul}
\usepackage[super]{nth}

\DeclareSIUnit{\rydberg}{Ry}
\DeclareSIUnit{\Ang}{\angstrom}
\DeclareSIUnit{\G}{\text{\ensuremath{G_0}}}
\usepackage{xcite}

\usepackage{xr}
\makeatletter
\newcommand*{\addFileDependency}[1]{
  \typeout{(#1)}
  \@addtofilelist{#1}
  \IfFileExists{#1}{}{\typeout{No file #1.}}
}
\makeatother

\newcommand*{\myexternaldocument}[2][]{%
    \externaldocument[#1]{#2}%
    \addFileDependency{#2.tex}%
    \addFileDependency{#2.aux}%
}
\myexternaldocument[SI-]{supplementary}

\author{Nils Wittemeier}
\affiliation{Catalan Institute of Nanoscience and Nanotechnology - ICN2 (CSIC, BIST), Campus UAB, 08193 Bellaterra, Spain}

\author{Matthieu J. Verstraete}
\affiliation{NanoMat/Q-Mat/CESAM and ETSF, Universit{\'e} de Li{\`e}ge (B5) and ETSF, B-4000 Li{\`e}ge, Belgium}
\alsoaffiliation{Catalan Institute of Nanoscience and Nanotechnology - ICN2 (CSIC, BIST), Campus UAB, 08193 Bellaterra, Spain}

\author{Pablo Ordej\'on}
\affiliation{Catalan Institute of Nanoscience and Nanotechnology - ICN2 (CSIC, BIST), Campus UAB, 08193 Bellaterra, Spain}

\author{Zeila Zanolli}
\affiliation{Chemistry Department and, Debye Institute for Nanomaterials Science, Condensed Matter and Interfaces, Utrecht University and ETSF, PO Box 80.000, 3508 TA Utrecht, The Netherlands}
\alsoaffiliation{Catalan Institute of Nanoscience and Nanotechnology - ICN2 (CSIC, BIST), Campus UAB, 08193 Bellaterra, Spain}
\email{z.zanolli@uu.nl}

\title{Interference effects in one-dimensional moir\'e crystals.}

\begin{document}

\begin{abstract}
Interference effects in finite sections of one-dimensional moir\'e crystals are investigated using a Landauer-B\"uttiker formalism within the tight-binding approximation. We explain interlayer transport in double-wall carbon nanotubes and design a predictive model. 
Wave function interference is visible at the mesoscale: 
in the strong coupling regime, as a periodic modulation of quantum conductance and emergent localized states;
in the localized-insulating regime, as a suppression of interlayer transport, and oscillations of the density of states. 
These results could be exploited to design quantum electronic devices.
\end{abstract}

\maketitle

Quantum materials~\cite{Keimer2017,basov_towards_2017} are a class of materials that exhibit quantum effects at the macroscopic scale.
They offer the opportunity to realize paradigm-shifting quantum electronics. The controlled generation and manipulation of quantum states by electrical, magnetic, or optical means is a key challenge in bringing quantum materials to applications.
Twisted bilayer graphene (tBLG) is a prime example of a tunable quantum material. The twist angle, and the resulting bi-dimensional moir\'e pattern, controls the emerging material properties: 
for specific magic angles, strong interlayer coupling and flat bands arise, resulting in superconductivity and strongly correlated phases~\cite{Cao2018SuperCond,Cao2018CorrelatedInsulator}.
One-dimensional moir\'e systems are realized in double-wall carbon nanotubes (DWNT) and are determined by two parameters: the angle between the inner and outer tubes' chiral vectors (like tBLG) and the difference between their radii. 
These degrees of freedom control the effective interlayer interaction and the resulting electronic states. 
While the physics emerging in 2D moir\'e crystals has been studied extensively, the understanding of its one-dimensional counterpart is limited to the ideal infinite nanotube case~\cite{Koshino2015,bonnet2016charge,zhao2020observation} and the commensurate, telescopic nanotubes~\cite{Kim2002,Grace2004,Tamura2005,Uryu2005,Yan2005,Tunney2006}.
Koshino et al.~\cite{Koshino2015} devised a continuum model for the idealized infinite DWNT. They identify three regimes with unique electronic properties (localized insulating, strong and weak coupling) determined by the relative alignment of the chiral vectors of the tubes.
Dispersionless flat bands appear in the localized insulating regime. The electronic structure of DWNT with weak coupling is given by the superposition of the constituent nanotubes, whereas it is heavily perturbed in the strong coupling regime: semiconducting tube combinations can produce a finite density of states in their gap, and metallic tubes can become semiconducting.
Experimental evidence of DWNTs with strong interlayer coupling has been recently reported~\cite{zhao2020observation}, but the relation between the measurements and the idealized infinite DWNT model by Koshino is not trivial. Experimental conditions impose a finite tube length and electrical contacts on the outer tube only. As the overlap region between inner and outer tubes is finite, strong confinement effects can arise, as in SWNTs with finite length~\cite{Rubio1999,rochefort_effects_1999}.
It is not clear {\it a priori} whether the regimes will be visible, in what limits of nanotube overlap  they could be recovered, and whether the regime separation is even valid for finite tubes. Answering these basic questions is essential for the use of coaxial nanotubes in applications involving electronic transport\cite{Nakar2020}.

In this work, we address these questions by studying interlayer transmission between two concentric nanotubes that overlap in a finite region and extend infinitely in opposite directions 
(telescopic double wall nanotubes, tDWNT, Fig~\ref{fig1:10-10_15-15-bands+trans}a).
We predict the effects of the length of the overlap region in the different regimes, going beyond previous works that only consider commensurate tDWNT.
The inter-layer transmission in tDWNTs composed of two armchair nanotubes (armchair@armchair) oscillates as a function of the overlap length, giving rise to regions with zero and maximal transmission. The overlap length modulates the tDWNT electronic properties. 
We demonstrate that this tunability extends to strongly coupled, chiral and incommensurate tDWNts.
We explain that dips in the transmission spectra 
of armchair@armchair tDWNTs emerge due to back-scattering by localized states in the overlap region. 
These transmission dips are absent in strongly coupled chiral tDWNT due the lack of rotational symmetries.
We devise a predictive model for the transmission, based on wave interference in one dimension, that reproduces the tunability and position of transmission dips. 
%
In the localized insulating regime we show that the interlayer conductance is very small irrespective of the overlap length, in agreement with Ref.~\citenum{Uryu2004}. 
We recover the Koshino limit for sufficiently long (but finite) overlap lengths, and provide benchmarks for experimental realizations of 1D moir\'es and correlated states in DWNT.
In the weak coupling regime, we show that interlayer transmission for chiral or incommensurate tDWNTs is suppressed.
tDWNTs composed of metallic zigzag nanotubes (zigzag@zigzag) also belong to the weak coupling regime, but are an exception to this rule: the inter-layer transmission can be significant if states on the inner and outer tube with different angular momentum couple. This coupling is subject to selection rules based on the chiral indices of the tubes involved~\cite{Kim2002}.

\section{Results and Discussion}

\begin{figure}
    \centering
    \begin{minipage}{\linewidth}%
        \includegraphics[width=\linewidth]{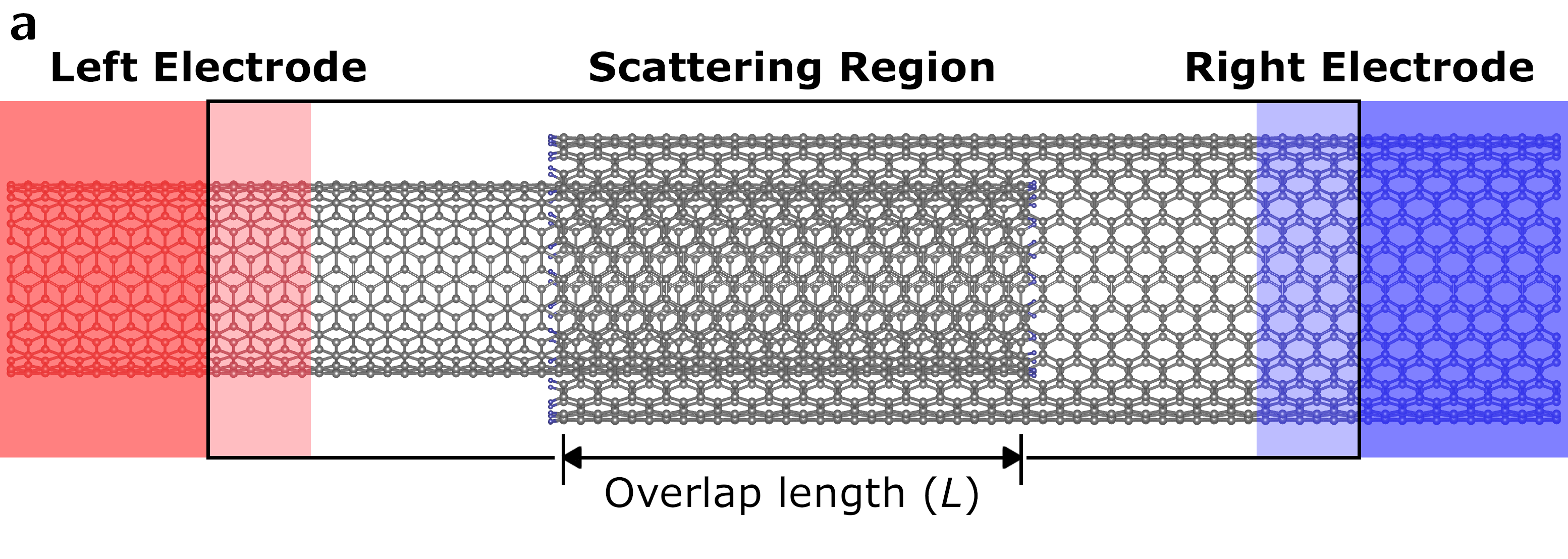}
    \end{minipage}
    \begin{minipage}[t]{0.49\linewidth}%
        \centering
        \includegraphics[width=\linewidth]{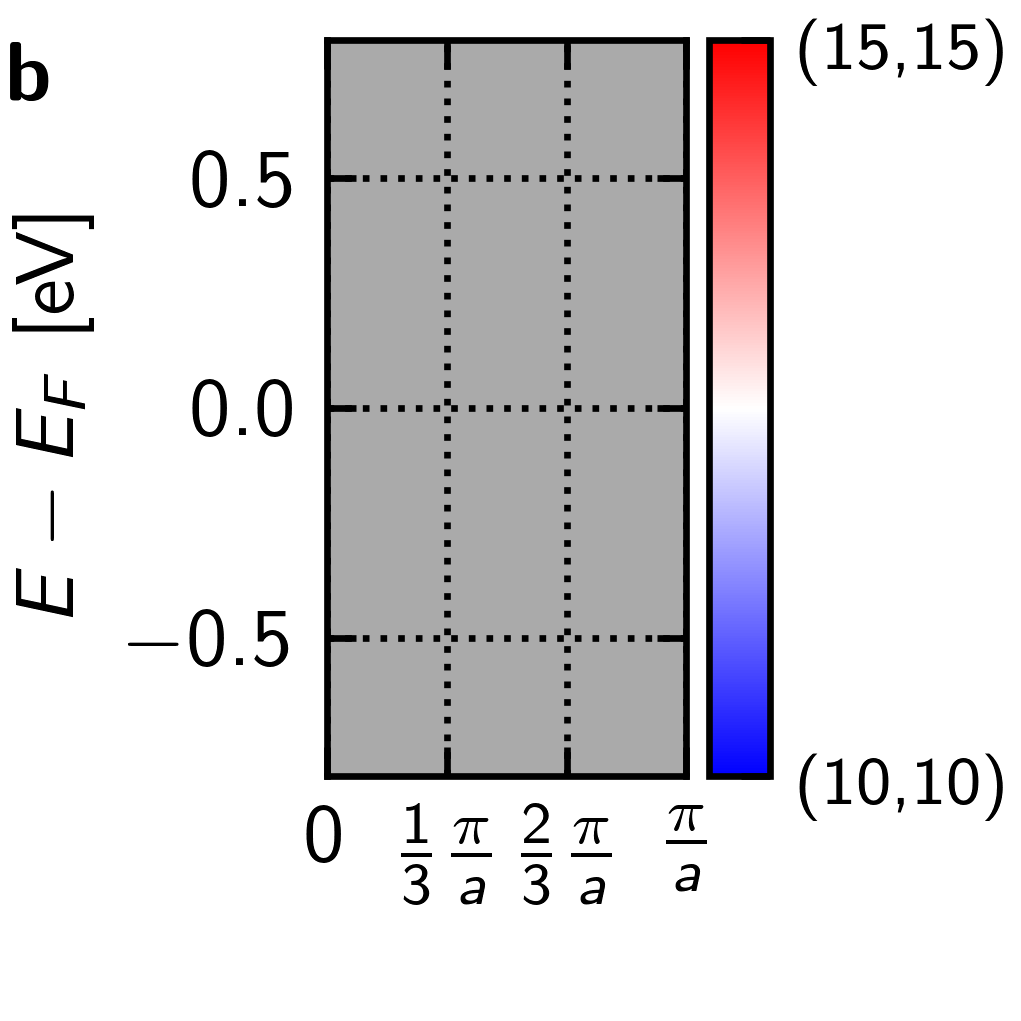}
    \end{minipage}%
    \begin{minipage}[t]{0.49\linewidth}
        \centering
        \includegraphics[width=\linewidth]{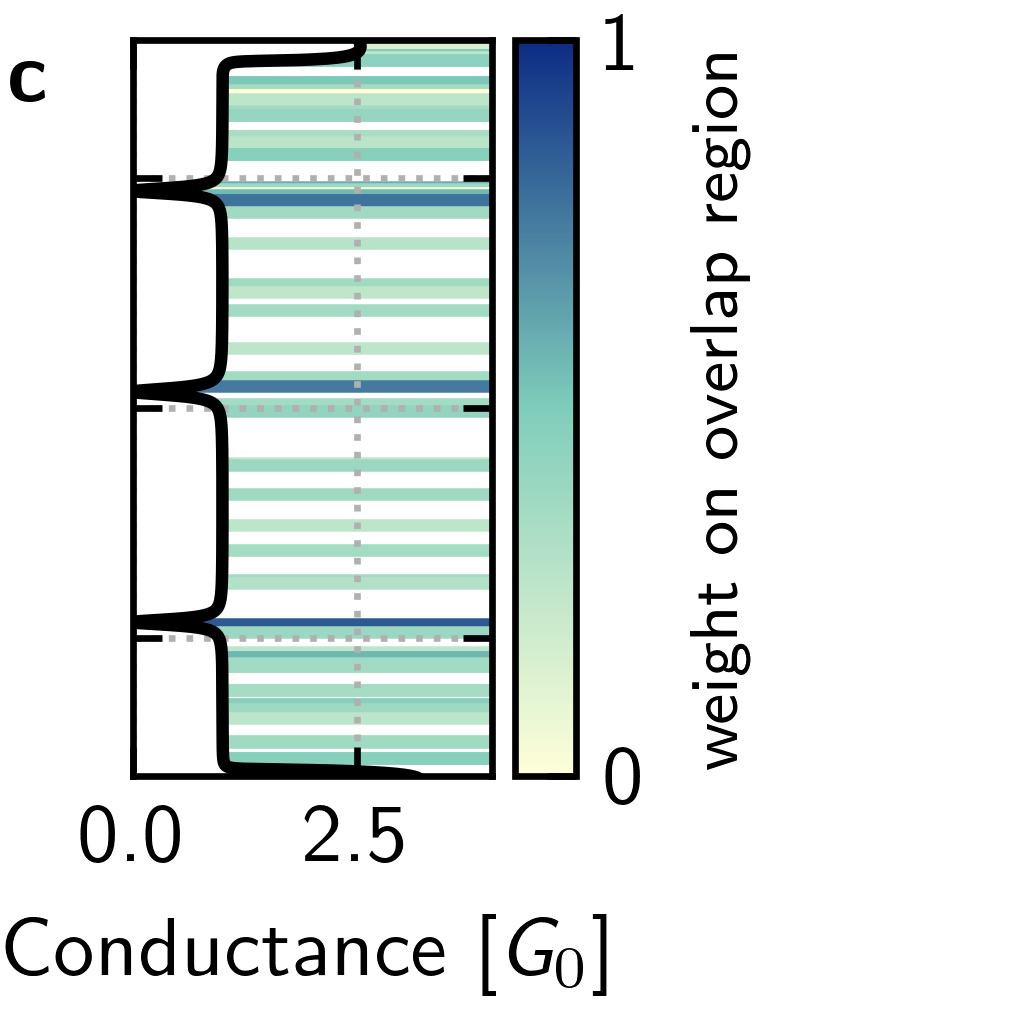}
    \end{minipage}
    \caption{%
    a) Transport set-up: 
    The scattering region consists of overlapping nanotubes, screening region, and part of the electrodes (light red/blue), which extend in opposite semi-infinite directions, and function as leads (red/blue).
    b) Projected TB band structure 
    of an ideal (10,10)@(15,15) DWNT.
    The bands with negative Fermi-velocity (gray) are hybridized between the nanotubes, bands with positive Fermi-velocity (red/blue) are localized on one tube each.
    c) TB electron transmission 
    from a semi-infinite (10,10) into a semi-infinite (15,15) SWNT through an overlap region with length \SI{34.08}{\Ang}.
    Horizontal lines visualize energy levels for an isolated, finite section of the system with additional \SI{44}{\Ang} of leads on either side. The yellow-to-blue scale indicates the weight of wave functions in the overlap region.
    Each dip in conductance coincides (approximately) with one or several localized eigenstates.
    \label{fig1:10-10_15-15-bands+trans}}
\end{figure}
%
Conductance simulations are performed for two concentric nanotubes that overlap over a finite length $L$, connected to semi-infinite SWNT leads (Fig.~\ref{fig1:10-10_15-15-bands+trans}a).
Electron transmission is only allowed at energies where conduction channels are available in both electrodes (only metallic nanotube contacts allow transmission close to the Fermi level). 
When electrodes consist of semi-conducting tubes, transmission can be achieved for chemical potentials within the valence or conduction bands of both nanotubes.
In either case, the total transmission $T$ through this asymmetrically contacted system can not exceed the smallest of the two electrode transmissions.
The magnitude of $T$ depends on how electrode states couple through the overlap region, {\it probing sensitively the interlayer interaction}.
%

\subsection{Strong coupling regime}
\begin{figure}
    \centering
    \includegraphics[width=\linewidth]{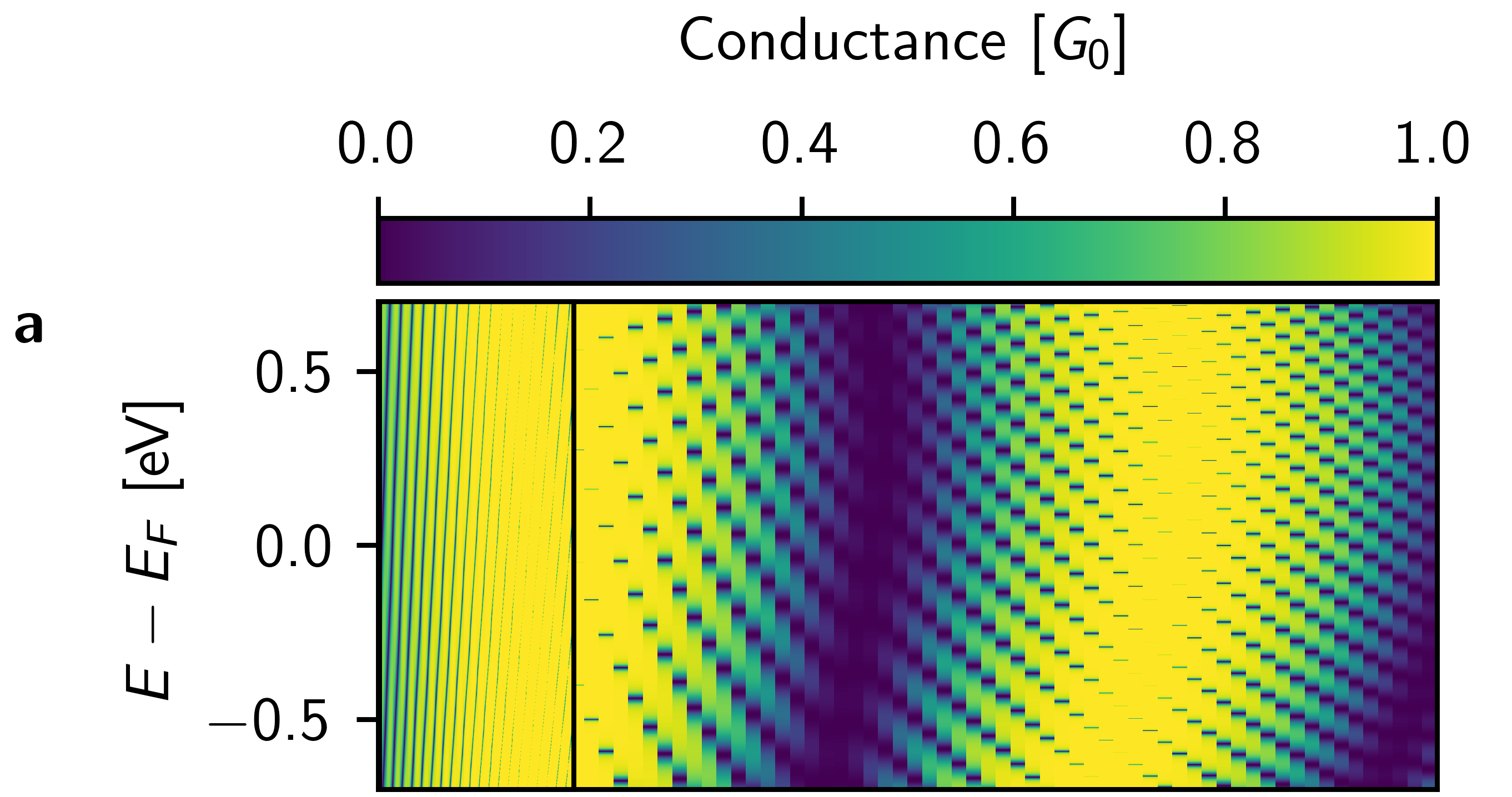}
    \includegraphics[width=\linewidth]{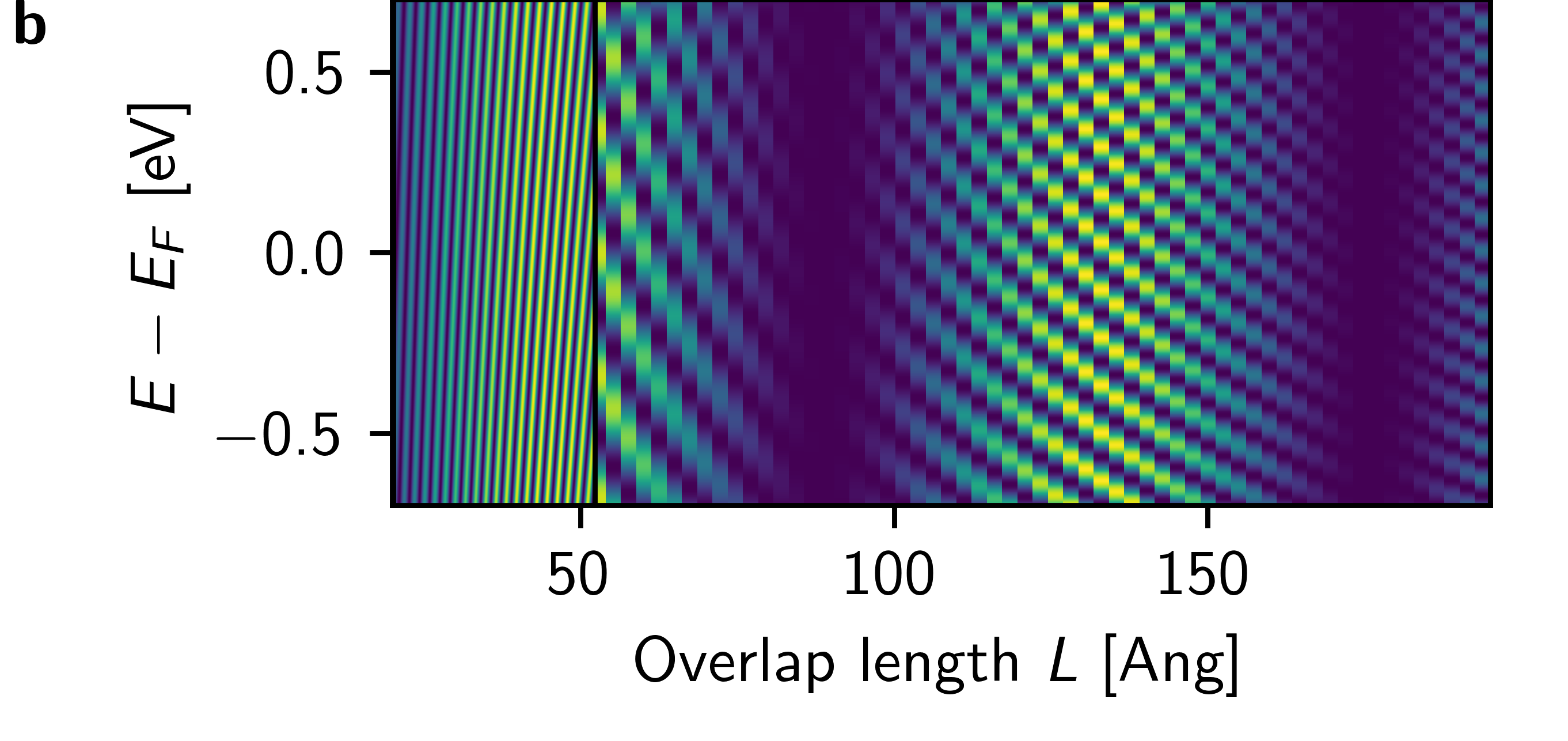}
    \caption{TB electron transmission from a (10,10) SWNT into a (15,15) SWNT through a finite overlap region at different overlap lengths calculated with \textit{a)} LB+TB formalism, and \textit{b)} our model (Eq.\ref{eq.model}). Overlaps are sampled with a high density below \SI{50}{\Ang} ($1/40\cdot\SI{2.459}{\Ang}$) and lower density above ($1/\SI{2.459}{\Ang}$).
    The model reproduces both trends in LB-transmission: the energy independent modulation of the transmission with overlap length on long spatial period, and the secondary modulation of the transmission dependent on energy and overlap.}
    \label{fig2:transmission-2D}
\end{figure}

DWNTs are in the strong coupling regime when the constituting tubes' chiral vectors are nearly parallel and their difference points along the armchair direction.
DWNTs consisting of armchair tubes (n,n)@(m,m) fulfill both conditions. 
Armchair single-wall nanotubes (SWNT) 
present two nearly linear bands crossing at the Fermi energy, resulting in a metallic system with \SI{2}{\G} conductance~\cite{Charlier2007,Zanolli2009}. Without interlayer interaction, the band structure and conductance of an ideal DWNT would simply be the sum of the individual SWNT ones.
Deviations from the ideal case can only occur for  perturbations of the sidewalls, for instance due to defects \cite{Zanolli10,ZanolliPRB09} or functionalization \cite{Zanolli12,Zanolli11}. 
The infinite (10,10)@(15,15) DWNT, instead, features (Fig.~\ref{fig1:10-10_15-15-bands+trans}b) a subtle band structure of states localized on either tube (positive Fermi velocity, red/blue) and hybridized states (negative Fermi velocity, gray). The conductance of the DWNT with finite overlap region shows a maximum of \SI{1}{\G} computed in the set-up of Fig.~\ref{fig1:10-10_15-15-bands+trans}a.
The transmission function presents a series of dips where conduction is blocked (Fig.~\ref{fig1:10-10_15-15-bands+trans}c). The number, position and width of the dips depend on $L$ (Fig.~\ref{fig2:transmission-2D}a). Each of these dips corresponds to a localized eigenstate in the overlap region (Fig.~\ref{fig1:10-10_15-15-bands+trans}c horizontal lines). 
At $L\approx\SI{50}{\Ang}$ and $L\approx\SI{150}{\Ang}$ the dips in transmission are barely visible. At intermediate overlap lengths, the interlayer conductance is reduced, the dips are broader, and transmission is blocked completely for $L\approx\SI{100}{\Ang}$ and
$L\approx\SI{200}{\Ang}$, following a periodic behavior.
This reduction of the maximum conduction as well as the oscillation with overlap length was predicted in Ref.~\citenum{Tamura2005}, and we can now show that localized states cause the dips observed in the conductance at fixed overlap lengths. We further show that the modulation of the transmission can be observed in a wide range of energy and overlap lengths.
%
%
%

The reduction of the maximum conductance near the Fermi energy with respect to the pristine DWNT can be explained taking into account the nature of the transport setup and the character of the bands in the DWNT. The asymmetric setup limits the conductance from \SI{4}{\G} to the minimum of the electrode conductances (\SI{2}{\G}). The reduction of the maximum conductance to \SI{1}{\G} is traced back to the three-fold rotational symmetry of the DWNT section, which results in the cancellation of one of the two available transport channels in Ref.~ \citenum{Tamura2005}. We note here that this reduction can already be inferred from the eigenstates of the pristine DWNT. The electronic eigenstates with positive Fermi-velocity are unaffected by the interlayer interaction, and have to scatter back into the electrode, effectively limiting the possible available channels to the electrode bands with negative Fermi velocity. These electrons couple to the two hybridized DWNT states, and have the possibility to be transmitted through the device. Depending on the overlap length this transmission occurs with almost no back-scattering ($L\approx\SI{50}{\Ang},\SI{150}{\Ang}$) or is completely blocked ($L\approx\SI{100}{\Ang}$).
In order to understand the origin of the transmission dips, we calculate the eigenstates of finite segments of the open system including additional CNT-rings on each side of the scattering region. In the limit of infinitely long electrode regions, we would recover the states of the open system. We find that each dip corresponds to a localized state (Fig.~\ref{fig1:10-10_15-15-bands+trans}c), and that the agreement between  eigenenergies and dip energies varies slightly, depending on the exact number of additional CNT-rings. The eigenstates localized in the overlap region ($L=\SI{34.08}{\Ang}$) are all similar in shape: the wave function weight on the atoms at the tube terminations are the highest, and they decay towards the electrode regions (SI Fig.~\ref{SI-fig:10-10_15-15-wfcs}).

\begin{figure}
    \centering
    \includegraphics[width=\linewidth]{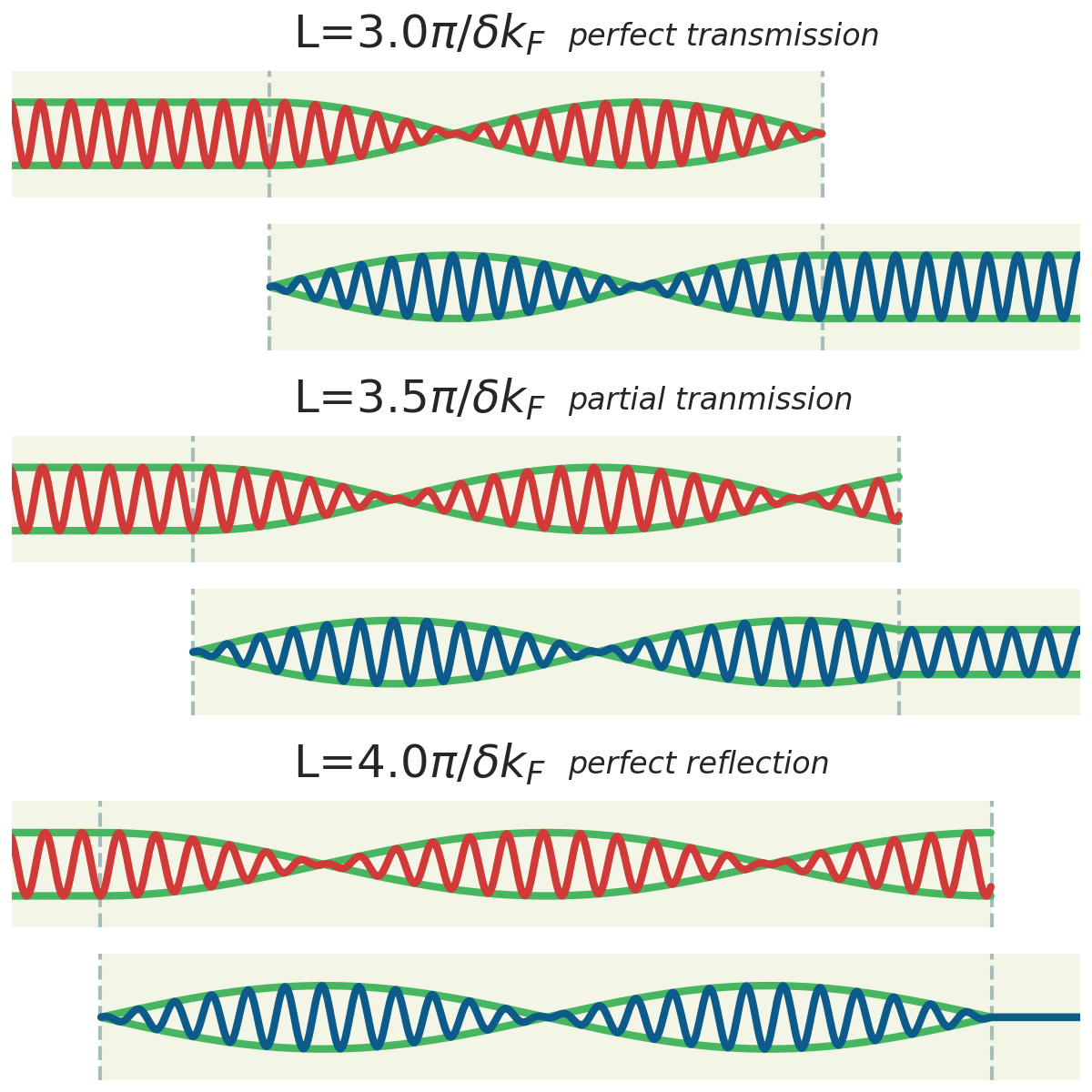}
    \caption{Schematic of standing waves in overlap region between two semi-infinite nanotubes.}
    \label{fig3:wave-scheme}
\end{figure}

To explain the origin of the localized states and the modulation of the transmission, we construct a simple model, assuming linear dispersion of the SWNT and DWNT bands with one common Fermi wave vector $k_F$ and velocity $v_F$. We further assume that the energy separation of the bands with negative $v_F$ in the DWNT is symmetric, giving two new vectors at each energy $k^{\pm}(E) = k(E) \pm \frac{1}{2}\delta k_F$. An incoming electron with energy $E= -v_F (k-k_F)$ needs to couple to the hybridized states with the same energy to pass from one layer into the other. The resulting superposition of overlap states propagates with two wave vector components: the average wave vector $k$, and the wave vector difference $\delta k_F/2$ modulating the incoming wave.
As this superposition propagates through the scattering device its weight oscillates between the two tubes. If the wavelength of the modulation is commensurate with the overlap region ($\delta k_F L= 2n\pi$) the incoming electron will be reflected at the termination and scatters back into the electrode. However, if the overlap length fits $(2n+1)/4$ wavelengths of the modulation ($L=(2n+1)\pi/2\delta k_F$) the electron passes through the overlap region without back-scattering (Fig.~\ref{fig3:wave-scheme}). At intermediate lengths, the incoming wave is partially transmitted reducing the conductance without fully blocking it. Similarly, at certain energies, which depend on the overlap length, the primary component $k$ becomes commensurate with the overlap. This allows a standing wave to form in the quantum box of the overlap region, which blocks the transmission and explains the emergence of dips at specific energies.
We combine the two trends to model the transmission using sine functions:
\begin{align}
    &T(E,L) = \sin(L\frac{\delta k_F}{2})^2
        \cdot \sin(L(k_F - \frac{E}{v_F}))^2 \label{eq.model} \\
    &T(E,\frac{2\pi n}{\delta k_F}) 
        = 0\nonumber\\
    &T(-v_F(\frac{\pi n}{L}+k_F),L) 
        = 0\nonumber
\end{align}
Fig.~\ref{fig2:transmission-2D} shows the impressive agreement between the transmission calculated with this simple model and the tight-binding results.
The model reproduces the energy-independent global oscillation of the transmission with the overlap length, and allows us to predict the periodicity of the transmission oscillation from the shift of the Fermi vector ($\delta k_F\approx\SI{0.071}{\per\Ang}$) for any nanotube combination.
For this tube shift ($\delta k_F \approx \SI{0.065}{\per\Ang}$) the corresponding periodicity is \SI{89}{\Ang}. The periodicity observed in our simulations is only slightly larger (\SI{100}{\Ang}). The curvature of regions without transmission (dark blue) can be recovered when the non-linearity of the bands is included in the model (Supplementary information). The model also reproduces the $E$ and $L$ dependent position of the smaller dips observed in the transmission. Furthermore, the shape of localized eigenstates matches the expected character (Fig.~\ref{fig3:wave-scheme}). The sine functions make the dips smoother compared to the LB-calculation, but the general trends are accurately reproduced. In fact this simple expression is an approximations of the more accurate one derived in Ref.~\citenum{Kim2002} when the linear-band approximation is applied. Numerical interpolation of the band structure allows for very accurate reproduction of the LB+TB transmission (SI Fig.~\ref{SI-fig:10-10_15-15-trans+accurate}). However, the simple approach is qualitatively correct and can intuitively be connected to the image of 1D-waves. 

\begin{figure}
    \centering
    \includegraphics[width=0.57\linewidth]{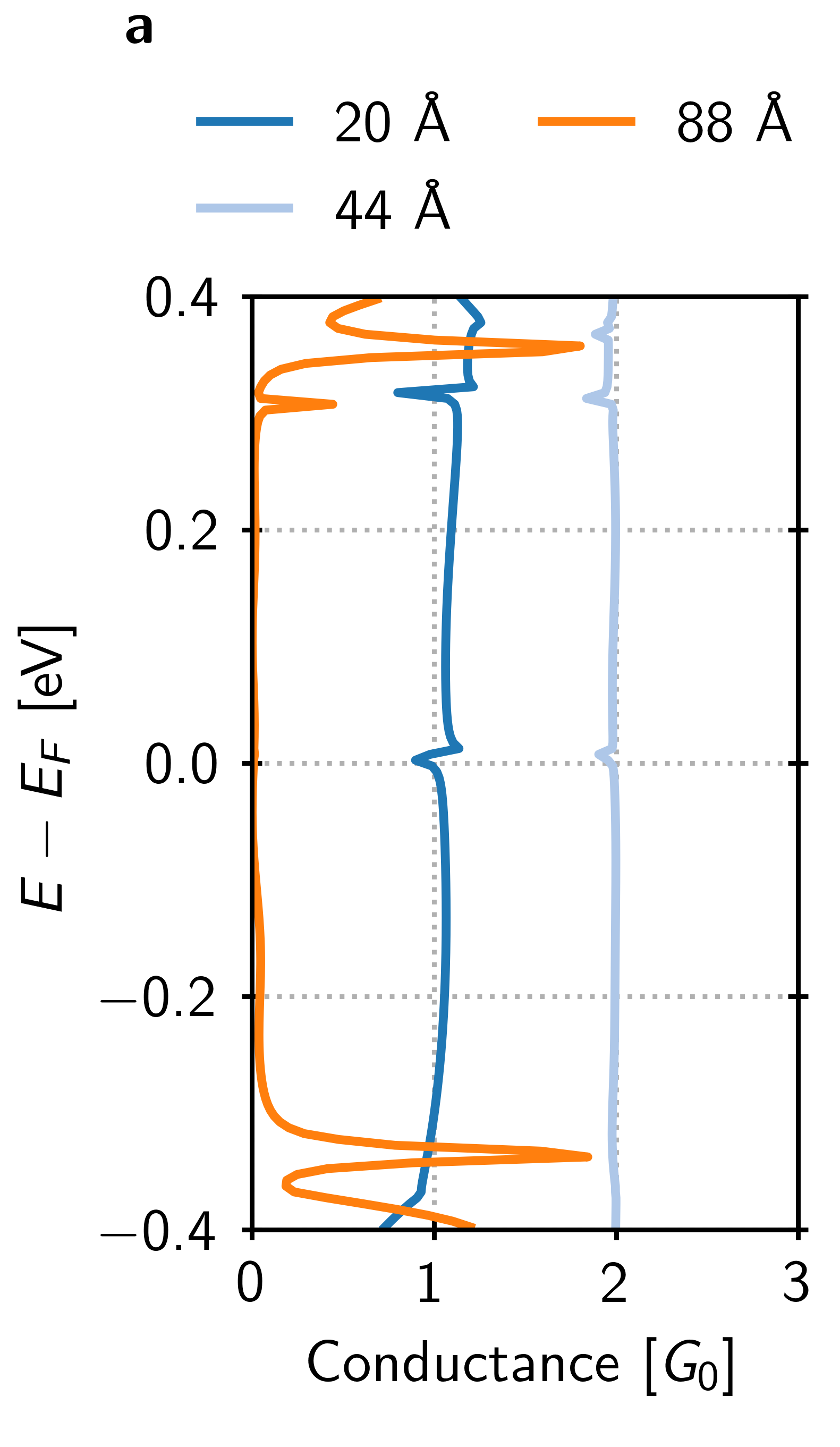}
    \includegraphics[width=0.41\linewidth]{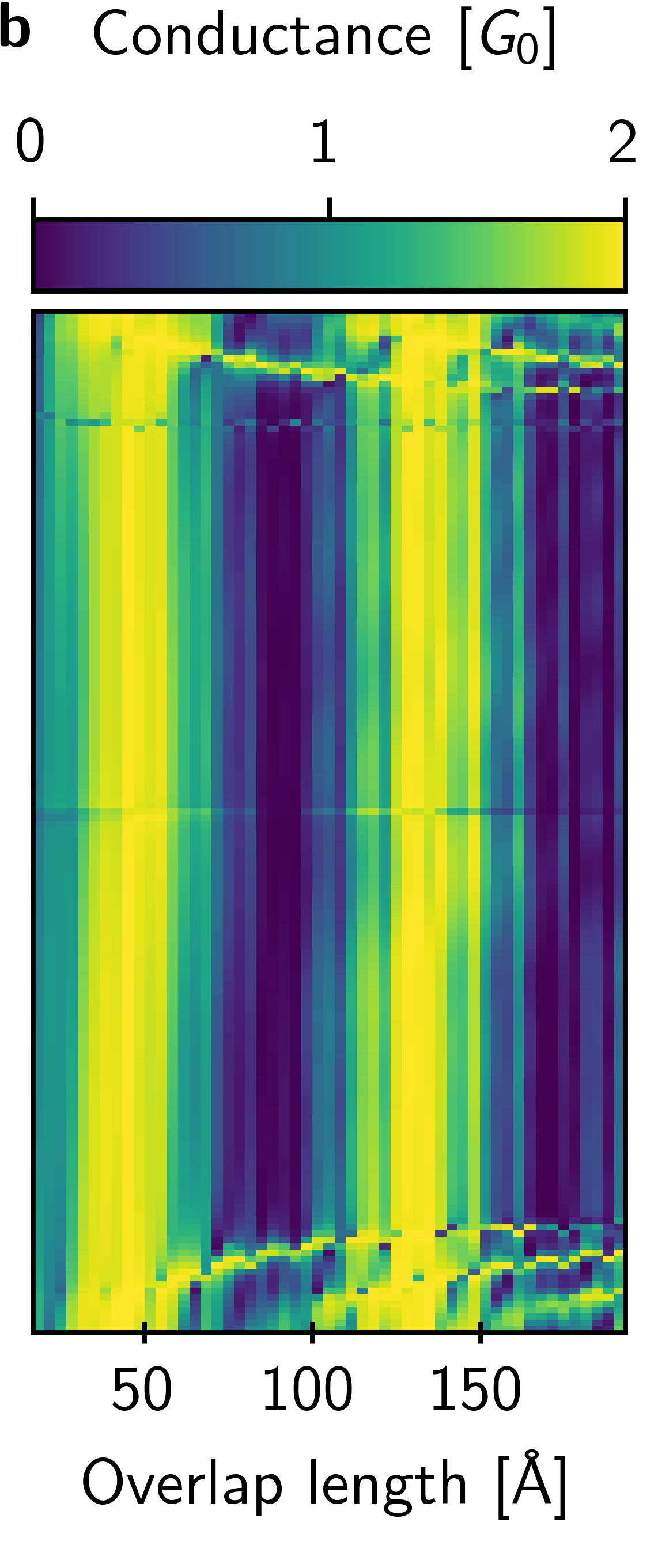}
    \caption{
    Strong coupling (M@M): 
    TB electron transmission from (18,15) into (23,20) with different overlap lengths $L$. The transmission oscillates as function of $L$. At \SI{0}{\eV} and \SI{0.31}{\eV} sharp features occur irrespective of $L$.}
    \label{fig:18-15_23-20-trans}
\end{figure}

We now consider an example of an {\it incommensurate} and chiral, strongly coupled, metallic DWNT: (18,15)@(23,20). The band structure can be computed by artificially imposing periodicity using a commensurate supercell consisting of 4 and 3 repetitions of (18,15) and (23,20), respectively, and straining both tubes by $\pm1\%$ 
(SI Fig~\ref{SI-fig:18-15_23-20-bands}). For the telescopic setup, the maximum conductance ($\SI{2}{\G}$) can be attained due to the strong hybridization of both linear bands between the tubes, resulting in two simultaneously available conduction channels.
The maximum of the conductance near $E_F$ (Fig~\ref{fig:18-15_23-20-trans}a) oscillates with the overlap length between $\SI{2}{\G}$ and $0$, with a period of \SI{90(1)}{\Ang} (Fig~\ref{fig:18-15_23-20-trans}b). Our simple model of wave function interference in a quantum box predicts a spatial period of \SI{96}{\Ang}, using a shift of the Fermi wave vectors $\delta k_F \approx \SI{0.065}{\per\Ang}$ extracted from the band structure in~\cite{Koshino2015}.
Despite its simplicity, our model is semi-quantitative and catches the main physical mechanisms at play: the oscillations of the transmission are caused by the difference in the Fermi wave vectors of the two tubes, and will appear in all combinations of strong coupling, metallic@metallic nanotubes.
The suppression of transmission at discrete energies observed in (10,10)@(15,15) results from achirality, and does not occur in chiral nanotubes. The wave functions of chiral tubes possess no rotational symmetry around the tube axis. Therefore they can only have nodes at few, specific points along the tube circumference, and cannot form localized states in the overlap region (SI Fig~\ref{SI-fig:18-15-wfc-2181}). 
Avoided crossings in the band structure~\cite{Koshino2015} result in sharp dips at \SI{0}{\eV} and \SI{0.31}{\eV} when the conductance is substantial ($L\approx\SI{44}{\Ang}$, $L\approx\SI{20}{\Ang}$), and peaks at ~$\SI{\pm0.31}{\eV}$ when conductance is suppressed ($L\approx\SI{88}{\Ang}$) (Fig.~\ref{fig:18-15_23-20-trans}).

\begin{figure}
    \centering
    \includegraphics[width=\linewidth]{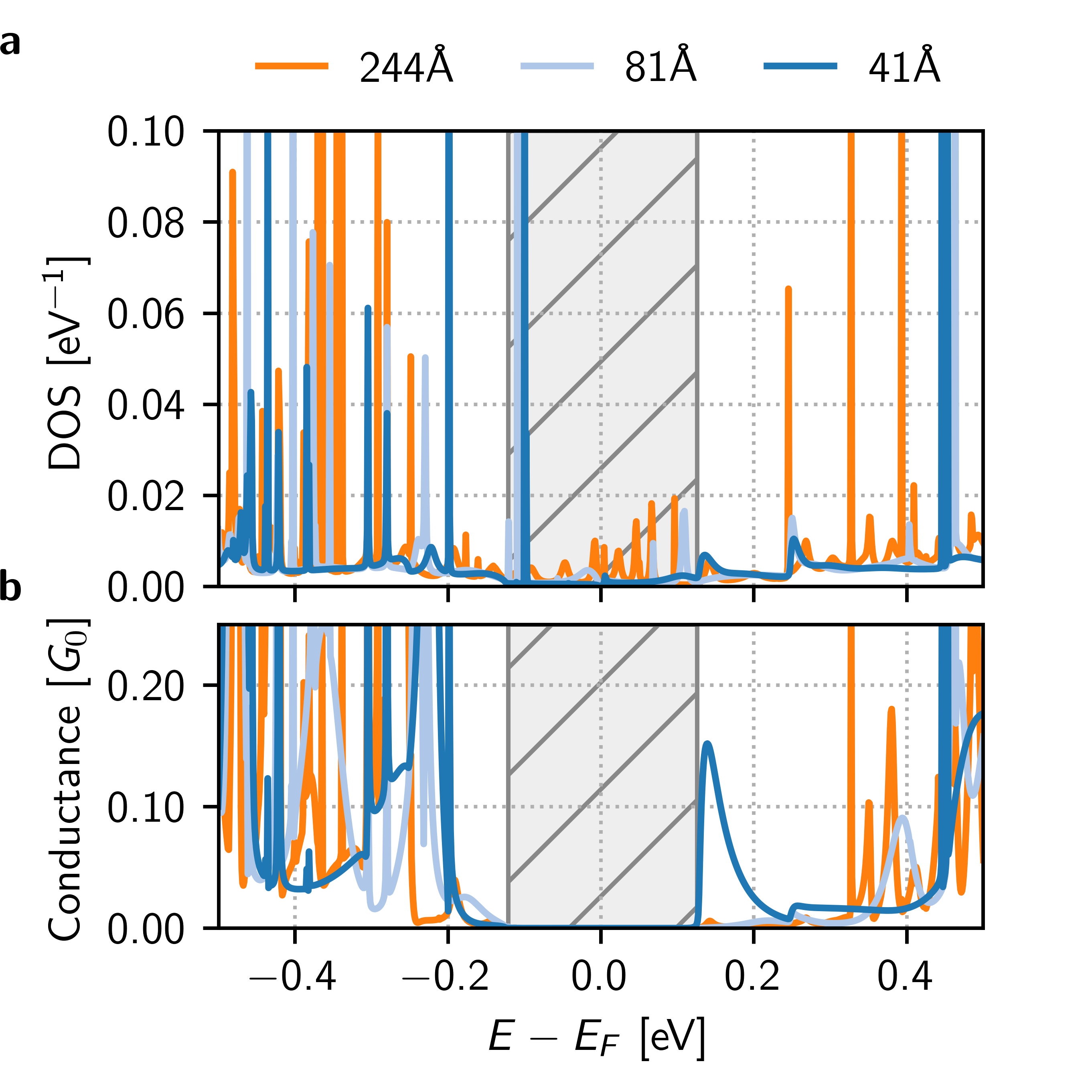}
    \caption{
    Localized insulating DWNT (M@SC): 
    density of states in the scattering region (a) and electron transmission (b) from (27,3) into (35,3) for different overlap lengths. 
    Interference between the tubes results in sharp spikes and flat bands (SI Fig.~\ref{SI-fig:27-03_35-03-bands}), which are more pronounced with increasing overlap length. The transport gap of the outer tube is highlighted in striped grey.
    \label{fig:27-03_35-03-tb-trans-dos}}
\end{figure}

\begin{figure}
    \centering
    \includegraphics[width=\linewidth]{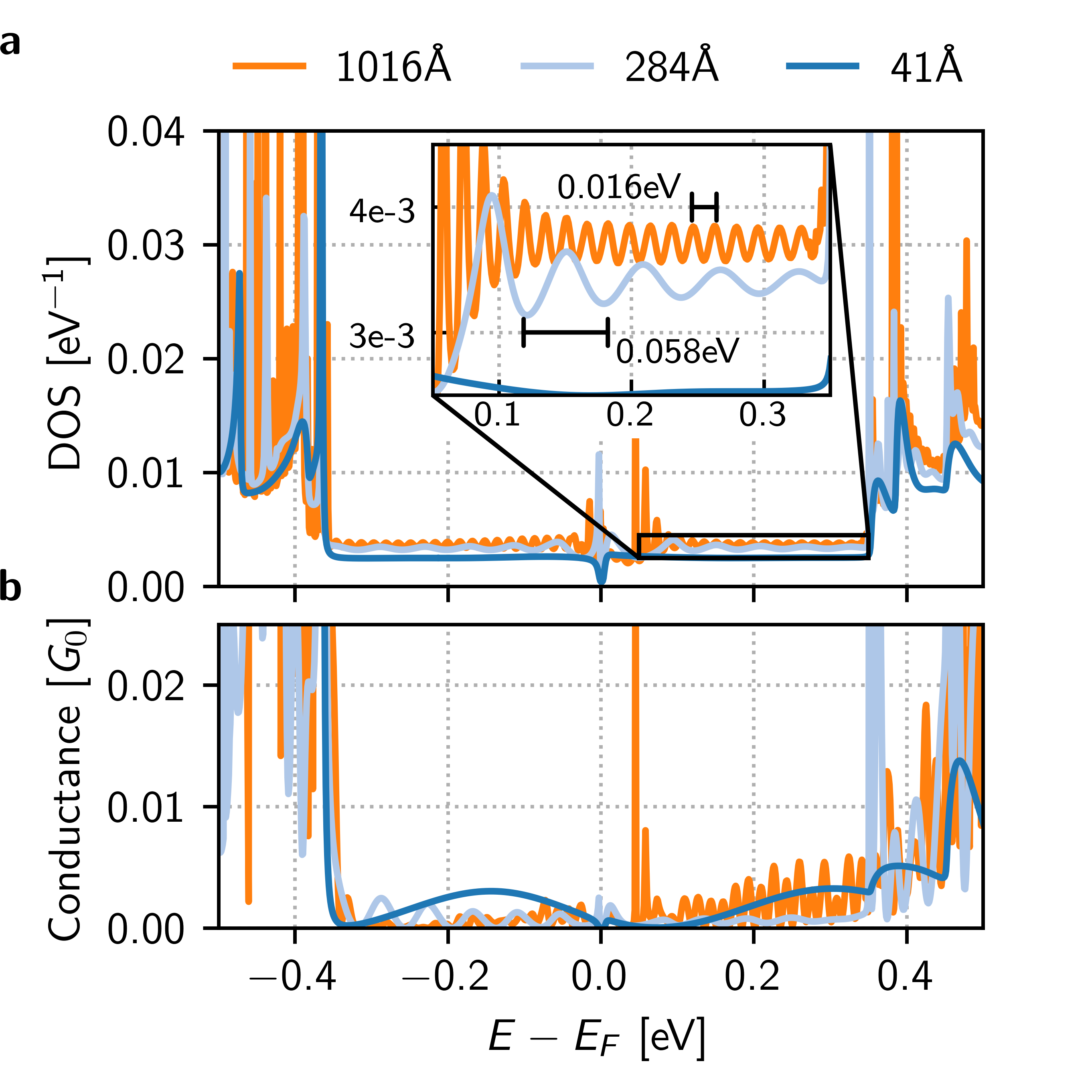}
    \caption{Localized quasi-insulating (M@M): density of states in the scattering region (\textbf{a}) and electron transmission (\textbf{b}) from (27,3) into (36,3) for different overlap lengths. 
    Interference between the tubes result in sharp spikes and flat bands 
    at $E_F$ and at the edge of the quasi-insulating gap, which are more pronounced with increasing overlap length.
    \label{fig:27-03_36-03-tb-trans-dos}}
\end{figure}

\subsection{Localized insulating regime}
The localized insulating regime 
is achieved when the chiral vectors of the inner and outer tube are nearly parallel and their difference points along the zig-zag direction. We consider two representatives of this family, (27,3)@(36,3) and (27,3)@(35,3), which are metallic-metallic (M@M) and metallic-semiconducting (M@SC), respectively. Interference effects are observable as peaks in the (27,3)@(35,3) DOS in Fig.~\ref{fig:27-03_35-03-tb-trans-dos}. The small (\SI{3.12}{\Ang}) interlayer spacing enhances the coupling
and suppresses the increase in DOS near the onset of the outer tube parabolic bands (gray area in Fig.~\ref{fig:27-03_35-03-tb-trans-dos} and SI Fig.~\ref{SI-fig:27-03_35-03-bands_zoom}).
This effect can be observed already in short DWNT segments ($L<$\SI{81}{\Ang}) and is not exclusive to  the infinite tubes of Ref.~\citenum{Koshino2015}. 
In (27,3)@(36,3), instead, the interference is weaker due to the larger interlayer spacing (\SI{3.51}{\Ang}) and the onset of the parabolic bands is clearly visible at \SI{\pm0.35}{\eV} (Fig.~\ref{fig:27-03_36-03-tb-trans-dos}). 
A series of peaks are observed in the DOS of both cases ~(Fig.~\ref{fig:27-03_35-03-tb-trans-dos} and \ref{fig:27-03_36-03-tb-trans-dos}).
This is consistent with the emergence of flat bands (SI Fig.~\ref{SI-fig:27-03_35-03-bands}) predicted by Koshino et al., who attribute them to the localization of electrons in an effective potential with long spatial period ($\approx\SI{1200}{\Ang}$ for these two cases). However, our simulations indicate that localized states already emerge in much shorter tube segments ($L\approx\SI{244}{\Ang}$), and could be experimentally observable even without requiring long, pristine DWNT samples. While some of the localized states emerge at these short overlap lengths, the peak density in the DOS increases with overlap length and is significantly lower than predicted by Koshino. This shows that the minima of effective potential that cause localization, emerge successively with increasing overlap. The full set localized states will be observable only if the overlap length is larger than the period of the effective potential.

We observe additional oscillations of lower magnitude in the DOS of (27,3)@(36,3) DWNT, which can not be attributed to flat bands. These oscillations occur in the energy ranges between \SIrange{-0.35}{-0.05}{\eV} and \SIrange{0.1}{0.35}{\eV} and are a result of the finite overlap length which causes a quantization of the wave vectors of states in the finite overlap region. 
Analogous to the discussion in the strong coupling regime, we assume linear dispersion of the low energy bands and determine the energy spacing of states commensurate with the overlap length:
\begin{equation}
\delta E = v_F \frac{\pi}{L} \approx \SI{5.6}{\eV\Ang}\frac{\pi}{L},\label{eq:dE}
\end{equation}
which matches the observed periods in the DOS ($\delta E_{DOS}$) quite well (Table~\ref{tab:27-03_35-03-dos_period}). 
\begin{table}[tb]
\setlength{\tabcolsep}{8pt}
\renewcommand{\arraystretch}{1.3}
\begin{tabular}{ r r r } 
      L [\si{\Ang}] & $\delta E_{DOS}[\si{\eV}]$ & $\delta E[\si{\eV}]$\\ \hline
     1016 & 0.016 & 0.017 \\
     528 & 0.031 & 0.033 \\
     284 & 0.058 & 0.062 \\ \hline
\end{tabular}
\caption{Oscillation periods in DOS of a (27,3)@(36,3) tDWNT in the energy ranges \SIrange{-0.35}{-0.05}{\eV} and \SIrange{0.1}{0.35}{\eV} for three different overlap length extracted from TB+LB calculations ($\delta E_{DOS}$) and calculated using Eq.~\ref{eq:dE} ($\delta E$).}
\label{tab:27-03_35-03-dos_period}
\end{table}
Given the small size of these oscillations, they will likely be challenging to observe experimentally. These oscillations are not predicted by Koshino, since they are caused by the finite length of the overlap. This quantization effect is absent in the (27,3)@(35,3) DWNT, because its outer tube is semiconducting and, therefore, only one set of linear bands is present.
In both   localized insulating DWNTs considered here, the interlayer conductance is very small ($<\SI{0.01}{\G}$) around the Fermi energy 
(\SI{-0.35}{\eV} to \SI{0.35}{\eV}) and only becomes  relevant in those regions where non-linear bands are present in the inner and outer nanotubes. This demonstrates that not only the metallic/semi-conducting nature of the individual tubes is important for the conductance between layers, but that the coupling regime also plays a significant role. Consequently, metallic shells in multi-wall carbon nanotubes are likely to contribute less to the overall conductance than previously expected due to the low interlayer conductance, unless the strong coupling condition is fulfilled. This explains the low interlayer conductivity reported in Ref.~\citenum{Uryu2005}. Different examples of localized-insulating tDWNT (metallic@metallic) with shorter interlayer spacing ($\dd{R}$) can be found in SI Fig.~\ref{SI-fig:45-18_54-18-trans} ($\dd{R}=\SI{3.4}{\Ang}$) and \ref{SI-fig:399-336_408-336-trans} ($\dd{R}= \SI{3.1}{\Ang}$) .

\subsection{Weak coupling regime}
In the weak coupling regime there is a distinction  between DWNTs composed of two zig-zag nanotubes (zig-zag@zig-zag) and all others. In zig-zag@zig-zag tDWNTs the rotational symmetry plays an important role, specifically, the three-fold rotational symmetry of (9,0)@(18,0) leads to significant interlayer transmission near the Fermi-level~\cite{Kim2002}. Similar to (10,10)@(15,15), the commensurability allows the formation of localized states in the overlap region and  causes the transmission to be blocked at the corresponding energies. For other tDWNTs in this regime, composed of either incommensurable or chiral nanotubes, the interlayer transmission is negligible as a result of weak coupling (SI Fig.~\ref{SI-fig:09-00_18-00-trans}-\ref{SI-fig:39-00_36-18-trans}).

\subsection{Conclusion}
In conclusion, we show that the two main coupling regimes predicted for double wall carbon nanotubes (strong coupling and localized insulating) are present already in finite DWNT segments. These regimes should be identifiable in experimental set-ups even when DWNT segments are short. 
%
For strongly coupled armchair@armchair metallic nanotubes, the interlayer transmission is significant and modulated by two phenomena. It is blocked by localized states at discrete energies for any given overlap region length, and it oscillates with the overlap length due to wave interference in the overlap region. The rotational symmetry of the tDWNT can reduce the maximal transmission, blocking transmission through one of the electrode channels completely, as seen for (10,10)@(15,15), or partially~\cite{Tamura2005}.
In other strongly coupled tDWNTs, where at least one nanotube is chiral, wave interference also lead to an oscillation of the maximal transmission. However, the absence of rotational symmetry prevents the formation of localized states and the corresponding back-scattering.
Based on the picture of 1D waves in a quantum box, we derive an expression for the interlayer transmission of strongly coupled nanotubes. The simplified model describes the modulation of transmission with excellent accuracy. 
In weakly coupled zigzag@zigzag metallic tDWNTs, the interlayer transmission is strongly affected by the rotational symmetry. This leads to significant transmission between the layers of (9,0)@(18,0), while the interlayer-transmission in (18,0)@(27,0) is negligible.
In other weak coupling tDWNTs, where at least one tube is chiral or the two tubes are incommensurate, the interlayer transmission is negligible.
Lastly, in the localized insulating regime the interlayer transmission is also heavily suppressed near the Fermi-level. In this regime the emergence of flat bands causes strong oscillations in the DOS. Notably, the flat bands already appear in segments significantly shorter than the predicted spatial periodicity of the underlying effective potential ($\approx\SI{1200}{\Ang}$ for tubes discussed above). The number of localized states is directly linked to the overlap length.
Our results highlight that  applications like nano-eletronic switches based on tDWNTs are very sensitive to their structure. Chiral tDWNTs are the most promising candidates for such applications. These tubes preserve the oscillating behavior found in all strongly coupled tDWNTs, while the absence of rotational symmetries prevents back-scattering at localized states.


\section{Methods}
\subsection{Tight-binding model}
We model inter- and intra-layer interactions using a non-orthogonal tight-binding (TB) model with one orbital per carbon site, 
based on 
Reich et al.~\cite{Reich2002}, Laissardiere et al.~\cite{Laissardiere2010}, and Bonnet et al.~\cite{bonnet2016charge}.
We extend the model to include curvature effects, using pp$\sigma$ interactions in addition to the pp$\pi$ interactions, and 
interlayer hopping to third nearest neighbors. The interlayer interaction ranges up to a cut-off distance of \SI{5}{\Ang} and is described by exponential decays:
\begin{align}
    H_{pp\sigma/\pi}(d) &= H^{inter}_{pp\sigma/\pi} e^{\frac{(d-a_0)}{g_{pp\sigma/\pi}}} \\
    S_{pp\sigma/\pi}(d) &= S^{inter}_{pp\sigma/\pi} e^{\frac{(d-a_0)}{h_{pp\sigma/\pi}}} , 
\end{align}
where $a_0=\SI{3.35}{\angstrom}$ is the interlayer spacing of bulk graphite.
The TB parameters were fit to electronic band structures of single and double-wall carbon nanotubes obtained from first-principles density functional theory. We first fix the pp$\pi$ parameters of the intra-layer interaction by fitting to the band structure of graphene, to reproduce correct behavior in the limit of very large nanotubes. Next, we fix the pp$\sigma$ parameters of the intra-layer interaction to reproduce the \textit{ab initio} bands of nanotubes with radii between \SIrange{5.2}{6.2}{\Ang}: (9,9), (16,0) and (9,6) (SI-Fig. \ref{SI-fig:fit}). Lastly, we optimize the the interlayer parameters to reproduce the \textit{ab initio} bands of two DWNTs: (16,0)@(24,0) and (9,6)@(15,10). The optimized parameters are summarized in Table~\ref{tab:fit_param}.
The transmission and density of states (DOS) are extracted using the 
Landauer-B\"uttiker formalism (LB)
formalism~\cite{datta1995} as implemented in TBTrans~\cite{Papior2017}. All calculations are performed at zero voltage between the two electrodes.

\begin{table*}[htb]
    \centering
    \caption{Optimized tight-binding parameters.}
    \setlength{\tabcolsep}{8pt} 
\renewcommand{\arraystretch}{1.3} 
\begin{tabular}{%
l S[table-format=1.3]%
l S[table-format=1.3] S[table-format=1.3]%
l S[table-format=1.3] S[table-format=1.3]}
\hline
    \multicolumn{2}{c}{onsite}  & \multicolumn{3}{c }{intra-layer} &\multicolumn{3}{c }{inter-layer} \\
                    &   &   &  \multicolumn{1}{c}{$pp\sigma$} & \multicolumn{1}{c }{$pp\pi$} &  & \multicolumn{1}{c}{$pp\sigma$} &  \multicolumn{1}{c  }{$pp\pi$} \\
\hline
    $E_p$[\si{\eV}] & -2.04 & $H^{1nn}$[\si{\eV}] &  3.93      & -2.81    & $H^{inter}$[\si{\eV}] &  0.505     &  0.709   \\
                    &       & $S^{1nn}$           &  0.573     &  0.301   & $S^{inter}$           &  0.003     &  0.062   \\
                    &       & $H^{2nn}$[\si{\eV}] &  1.17      & -0.679   & $g$[\si{\angstrom}]   &  0.408     &  0.387   \\
                    &       & $S^{2nn}$           &  0.018     &  0.047   & $h$[\si{\angstrom}]   &  1.12      &  0.620   \\
                    &       & $H^{3nn}$[\si{\eV}] &  1.11      & -0.298   &                             &            &          \\
                    &       & $S^{3nn}$           &  0.074     &  0.040   &                             &            &          \\
\hline
\end{tabular}
    \label{tab:fit_param}
\end{table*}

\subsection{First principles calculations}
First principles calculations were done with the \textsc{siesta}~\cite{Soler2002,Junquera2020} implementation of the density functional theory (DFT) method. We employed the local density approximation (LDA) exchange-correlation functional as parametrized J. P. Perdew and Y. Wang~\cite{Perdew1992} in conjunction with optimized norm-conserving Vanderbilt (ONCV) scalar-relativistic pseudopotentials~\cite{Hamann2013,vanSetten2018} with $2s$ and $2p$ valence electrons. The Brillouin zone was sampled using a $\Gamma$-centered, one-dimensional grid with 78 k-points for the pristine armchair nanotubes.


\section{Acknowledgments}
The authors acknowledge enriching discussions on the tight-binding parametrization with L. Henrard and P. Lambin. 
This work was supported by Spanish MINECO (the Severo Ochoa Centers of Excellence Program under Grant No. SEV- 2017-0706), Spanish MICIU, AEI and EU FEDER (Grants No. PGC2018-096955-B-C4), Generalitat de Catalunya (Grant No. 2017SGR1506 and the CERCA Programme), and the European Union MaX Center of Excellence (EU-H2020 Grant No. 824143).
ZZ acknowledges financial support by the Ramon y Cajal program RYC-2016-19344 (MINECO/AEI/FSE, UE) and the Netherlands Sector Plan program 2019-2023.
MJV acknowledges funding by the Belgian FNRS (PDR G.A. T.1077.15-1/7 and a sabbatical ``OUT'' grant at ICN2), and computational resources from the Consortium des Equipements de Calcul Intensif (CECI, FRS-FNRS G.A. 2.5020.11) and Zenobe/CENAERO funded by the Walloon Region under G.A. 1117545.
NW has received funding from the European Union’s Horizon 2020 research and innovation programme under the Marie Skłodowska-Curie grant agreement No. 754558.
We acknowledge computer resources at MareNostrum4 at Barcelona Supercomputing Center (BSC), provided through the PRACE Project Access (OptoSpin project 2020225411) and RES (activity FI-2020-1-0014), and technical support provided by the Barcelona Supercomputing Center.
The authors acknowledge use of the open-source project sisl\cite{sisl2019} used to generate atomic structures and as basis of implementation for our tight-binding model. 


\bibliography{main}

\end{document}